\documentclass[%
 aps,
 prl,%
 amsmath,amssymb,superscriptaddress,
 preprint,%
]{revtex4-1}

\usepackage{graphicx}
\usepackage{bm}

\begin{document}

\title{Phonon nanocapacitor for storage and lasing of terahertz lattice waves}

\author{Haoxue Han}
\affiliation{CNRS, UPR 288 Laboratoire d'Energ\'etique Mol\'eculaire et Macroscopique, Combustion (EM2C), Grande Voie des Vignes, 92295 Ch\^atenay-Malabry, France}
\affiliation{Ecole Centrale Paris, Grande Voie des Vignes, 92295 Ch\^atenay-Malabry, France}%

\author{Baowen Li}
\affiliation{Department of Physics, Centre for Computational Science and Engineering, and Graphene Research Center, National University of Singapore, 117546 Singapore}%
\affiliation{NUS Graduate School for Integrative Sciences and Engineering, National University of Singapore, 117542 Singapore}%
\affiliation{Center for Phononics and Thermal Energy Science, School of Physics Science and Engineering,
Tongji University, 200092 Shanghai, P. R. China}%

\author{Sebastian Volz}
\affiliation{CNRS, UPR 288 Laboratoire d'Energ\'etique Mol\'eculaire et Macroscopique, Combustion (EM2C), Grande Voie des Vignes, 92295 Ch\^atenay-Malabry, France}
\affiliation{Ecole Centrale Paris, Grande Voie des Vignes, 92295 Ch\^atenay-Malabry, France}%

\author{Yuriy A. Kosevich}%
\affiliation{CNRS, UPR 288 Laboratoire d'Energ\'etique Mol\'eculaire et Macroscopique, Combustion (EM2C), Grande Voie des Vignes, 92295 Ch\^atenay-Malabry, France}
\affiliation{Ecole Centrale Paris, Grande Voie des Vignes, 92295 Ch\^atenay-Malabry, France}%
\affiliation{
Semenov Institute of Chemical Physics, \\ Russian Academy of Sciences, Kosygin str. 4, Moscow 119991, Russia
}%

\date{\today}

\begin{abstract}
We introduce a novel ultra-compact nanocapacitor of coherent phonons formed by high-finesse interference mirrors based on atomic-scale semiconductor
metamaterials.
Our molecular dynamics simulations show that
the nanocapacitor stores THz monochromatic lattice waves, which can be used for phonon lasing - the emission of coherent phonons.
Either one- or two-color phonon lasing can be realized  depending on the geometry of the nanodevice.
The two color regimes of the capacitor originates from the distinct transmittance dependance on the phonon wave packet incident angle
for the two phonon polarizations at their respective resonances.
Phonon nanocapacitor can be charged by cooling the sample equilibrated at room temperature or by the pump-probe technique.
The nanocapacitor can be discharged by applying tunable reversible strain, resulting in the emission of
coherent THz acoustic beams.
\end{abstract}

\maketitle

Phonons, quanta of lattice waves, having significantly shorter wavelengths than photons at the same frequency,
may enable us to pursue improved resolution in tomographic, ultrasonic and other imaging techniques using focused
sound waves.
  In particular, the terahertz (THz) phonons that have wavelengths comparable to the lattice constants would
  allow us to detect microscopic structure of the material up to the atomic scale with high precision.
  To this end, coherent phonons are urgently desired.
  Coherent lattice waves have been studied in semiconductor
superlattices\cite{huynh2006,lanzil2010,beard2010,maryam},
optomechanical systems\cite{kippenberg2008,grud2010,khurgin2012}
and electromechanical resonators\cite{mahboob2012,okamoto2013,mahboob2013}.
  Coherent phonon manipulations for energy transport in dynamical nanosystems
has been an emerging focus of research\cite{li2012}.
  The development of nanophononic devices which enable controllabe generation and coherent emission of
phonons has recently emerged as the subject of intense interest\cite{trigo2002,huynh2006,
lanzil2010,beard2010,grud2010,mahboob2013,fainstein2013,kippenberg2008,khurgin2012}.
  Conventional sources of sound waves, such as piezoelectric transducers, fail to operate efficiently above
a few tens of gigahertz (GHz).
The highest emitted-phonon frequency was achieved in a semiconductor superlattice under electrical
bias in which half-THz coherent phonons were released during electrons tunneling through the
quantum wells\cite{beard2010}.
  Similar coherent phonon emission was also observed
in hybrid optomechanical schemes in which optically pumped microcavities produce stimulated acoustic
emission in the range of MHz\cite{grud2010} and GHz\cite{fainstein2013}.
An entirely mechanical phonon laser at sub-MHz was achieved in an electromechanical resonator\cite{mahboob2012, okamoto2013}
in analogy with
a three-level laser scheme\cite{mahboob2013}.
For these systems, the need for strong optical pumping or complex actuator tuning may limit the development
of phonon lasers because unavoidable compromises have to be made in the design scheme.


In this Letter, we introduce a three-dimensional (3D) phonon nanocapacitor based on atomic-scale semiconductor metamaterials.
  The nanocapacitor allows for generation of THz coherent phonons with an ultra-high
monochromatic quality by an adiabatic cooling of the nanodevice or by the pump-probe optical technique \cite{huynh2006,lanzil2010}.
   We show that the nanocavity structure formed by high-finesse phonon interference mirrors (PIMs) can efficiently
store a large number of coherent out-of-equilibrium phonons.
  The nanocapacitor can
  emit coherent THz phonon beams upon application of tunable reversible strain. Such emission can be considered as ``phonon lasing''.
  Either one- or two-color phonon lasing can be realized depending on the geometry of the nanodevice, in contrast to the usual one-color photon lasing.
  The achievement of generation and emission of coherent THz lattice waves will provide an essential step towards
active hypersound devices and nanophononic applications of THz acoustics, including surgery with focused
ultra- and hyper-sound in medicine\cite{khokh2014}.


\begin{figure}
\includegraphics[width=8.0cm]{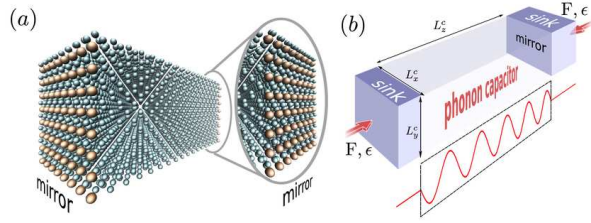}
\caption{\label{fig1}
\textbf{Silicon phonon capacitor composed of two phonon interference mirrors (PIMs) separated by a spacer.}
\textbf{a,} The atomistic view of the phonon capacitor and mirrors.
The atoms in brown are Ge impurity atoms and the green ones are the Si atoms of the host lattice.
\textbf{b,} Two heat sinks coupled to the phonon capacitor.
The dimensions of the capacitor are chosen as $L_x^c=L_y^c=8$ nm and $L_z^c=35$ nm.
}
\end{figure}
\begin{figure}
\includegraphics[width=8.0cm]{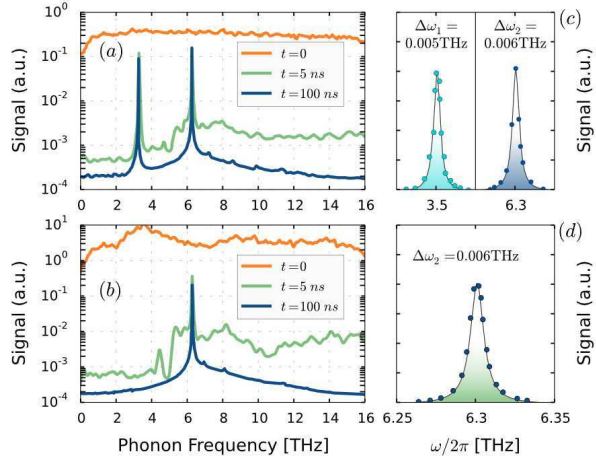}
\caption{\label{fig2}
\textbf{Power spectral density of atomic kinetic energy as a function of time delay $t$
after the cooling onset in different geometries.}
\textbf{a,b,} Nanocapacitor with $L_x^c=L_y^c=8$ nm and $L_z^c=35$ nm (\textbf{a}) and $L_x^c=L_y^c=33$ nm and $L_z^c=35$ nm (\textbf{b}).
\textbf{c,d,} Gaussian fit of the capacitor phonon peak to measure the full width at half maximum (FWHM) $\Delta\omega$
for the peaks in (\textbf{a}) and (\textbf{b}), shown by (\textbf{c}) and (\textbf{d}), respectively. The lines and symbols are Gaussian functions and MD results.
}
\end{figure}

  A detailed atomistic presentation of the phonon capacitor is depicted in Fig.~\ref{fig1}(a).
The capacitor consists of two PIMs separated by a spacer made of bulk silicon (Si).
The PIM is composed of an atomic-scale metafilm: an internal atomic plane
in Si lattice embedded with segregated Germanium (Ge) impurities atoms, as shown in Fig.~\ref{fig1}(a).
  The assisted laser Molecular Beam Epitaxy (MBE) have achieved atomically sharp interfaces in a superlattice
\cite{ravi2014}
and can hence provide a possible experimental implementation for our proposed nanocapacitor structure
since this technique can reach the resolution of a single unit cell in the lattice.
In the molecular dynamics (MD) simulations,
the covalent Si:Si/Ge:Ge/Si:Ge interactions are modeled by the Stillinger-Weber (SW) potential\cite{SW},
which includes both two- and three-body potential terms.
The SW potential has been widely used to study the thermal\cite{volz2001,mingo2004}
and mechanical\cite{atrash2011} properties
of Si and Ge materials for its best fit for experimental results.
  All the MD simulations were performed with\begin{footnotesize} LAMMPS \end{footnotesize}\cite{LAMMPS}.

  We first implement the charging of a phonon capacitor by cooling a sample equilibrated at
room temperature by MD simulations.
Two heat sinks are coupled to the capacitor on the two facets of the mirrors, as shown in
Fig.~\ref{fig1}(b).
The phonons in the capacitor are dissipated once entering the sinks,
which are modeled by a Langevin heat bath\cite{langevin}, yielding the effects of inelastic
scattering on phonon transport.
The capacitor was first thermalized homogeneously to $T=300$K to ensure that the material
approaches the state of energy equipartition. 
  The classical approximation remains relevant for Si at $T=300\text{K}\approx 1/2\Theta_D$, where $\Theta_D=645$K
is the Debye temperature of Si.
In Fig.~\ref{fig2}(a) and (b), the power spectral density of atomic kinetic energy in the capacitor is shown which was
calculated from the windowed Fourier transform to demonstrate that all phonon modes were excited.
Then the temperature of the heat sinks was set to 0K to cool down the capacitor.
Note that the switch-on of the heat sinks took only a few picoseconds, during which the phonons present in the
capacitor had no time to escape, which corresponds to an ``adiabatic cooling'' process\cite{sonntag1998}.
The idealized assumption of absolute zero temperature is considered to clarify the physical mechanisms.
In practice, liquid helium can be used for the heat sinks and Dewar flasks can be considered to
prevent heat conduction and radiation from the free surfaces in experimental implementations of the
phonon capacitor.

  We have found that the nanocapacitor can function in two regimes: in a dual-mode regime and in a single-mode
regime,  depending on the aspect ratio of the capacitor $p$ defined as $p=L_x/L_z=L_y/L_z$.
For a quasi-one-dimensional capacitor with small aspect ratio $p\approx 1/4$,
after $t=5$ns of cooling we notice a concentration of phonon energy at the frequencies $\omega_1=3.5$THz
and $\omega_2=6.31$THz, which corresponds to the dual-mode regime with the storage of both transverse and longitudinal coherent lattice waves.
The spectral energy for all phonon modes in the capacitor has decayed by over two
orders of magnitude except for the two modes, as shown in Fig.~\ref{fig2}(a) and (c).
By a cooling duration of $t=100$ns, which is very long for an atomic system,
the coherent phonon peak intensity at $\omega_{1,2}$ shows a decrease of
only $10\%$ as $t=5$ns while all the other phonon modes have practically escaped the capacitor.
  This high monochromatic quality makes the capacitor an ideal candidate for a
coherent phonon source - phonon laser\cite{grud2010,beard2010,maryam,fainstein2013}.


The single-mode regime was achieved in
a three-dimensional capacitor with $p\approx 1$: after the same cooling process, only the longitudinal mode at $\omega_2=6.31$THz is stored,
as shown in Fig.~\ref{fig2}(b) and (d).
The equivalent temperature $T_c$ of the non-equilibrium phonons in the capacitor can be estimated as 
$T_c=\hbar\omega_2/k_B=303$K $\gg T_e$, where $\hbar$ is the reduced Planck constant, $k_B$ is the Boltzmann
constant and $T_e$ is the surrounding temperature.
Since the capacitor phonons are in a single-mode state,
the phonon number $n_C$ and the total elastic lattice energy $E_{el}$
satisfies $E_{el}=\hbar\omega_2(n_C+\frac{1}{2})$.
We thus obtain a phonon number $n_C\approx 57,000$ with $E_{el}=77.8$eV stored in the present capacitor.
In the limit of $n_C\gg 1$, $n_C$ and $\omega_2$
play, respectively, the role of the effective charge and potential of the phonon nanocapacitor.
For a sufficient large phonon number $n_C$, the classical equations of motion are valid for ensemble averaging
in the semiclassical limit\cite{gardiner2004,aspelmeyer2013}.

\begin{figure}
\includegraphics[width=8.5cm]{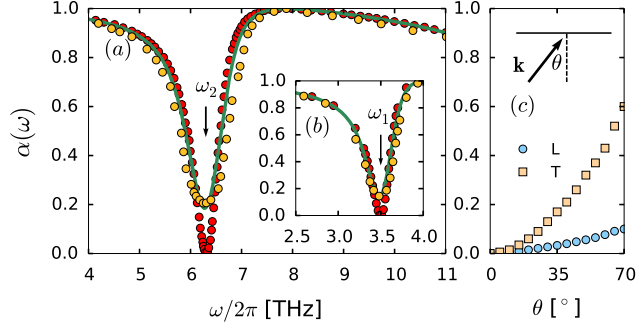}
\centering
\caption{\label{fig3}
\textbf{Spectra of the energy transmission coefficient through interference mirror obtained by MD simulations for phonon WPs with
different coherence lengths $\xi$ in Eq. (\ref{eqn1}) and incident angles $\theta$.}
\textbf{a,b,} Transmission of normally incident WP ($\theta=\frac{\pi}{2}$) with long coherence length $\xi=100\lambda_c$ (plane-wave approximation, red circles) and with short coherence length
$\xi=8\lambda_c$ (yellow circles) in comparison with the convolution (green curve) given by equation (\ref{eqn2})
for longitudinal (\textbf{a}) and transverse (\textbf{b}) WP polarization with respect to the mirror plane.
Under the plane-wave approximation $\xi\gg\lambda_c$ ($\Delta\omega\ll\omega_c$),
the transmission spectrum $\alpha_{\text{pw}}(\omega)$ displays
the  antiresonance profile marked by a sharp dip of zero transmission at $\omega=\omega_{1,2}$ due to the
two-path phonon destructive interference.
In the transmission of a narrow WP with $\xi=8\lambda_c$ and $\Delta\omega=0.06\omega_c$,
the interference effect is weakened, with $\alpha(\omega_{1,2})>0$,
by the presence of a large number of frequency components,
but strong reflection at the antiresonance remains pronounced.
\textbf{c,} Transmission of oblique incident WP ($\theta<\frac{\pi}{2}$) at the resonances $\omega=\omega_{1,2}$
for the longitudinal and transverse polarizations.
}
\end{figure}

The confinement of the phonon modes at $\omega_{1,2}$, stored in the capacitor,  results from the two-path phonon destructive
interference in the mirrors\cite{YAK1997,kosevich08,han2014}.
Ge atoms in the planar defect in the PIM force phonons to propagate through two paths: through unperturbed (matrix) and
perturbed (defect) interatomic bonds. The resulting phonon interference yields transmission antiresonance
(zero-transmission dip) in the spectrum of THz phonons.
By MD-based phonon wave-packet (WP) method\cite{han2014},
we calculated the energy transmission coefficient $\alpha(\omega,\textbf{k})$ for the Si:Ge interference mirror, for transverse and longitudinal waves.
To compute $\alpha(\omega,\textbf{k})$,
we excited a 3D Gaussian wave packet centered at the wave vector $\textbf{k}$ in the reciprocal space
and at $\textbf{r}_0$ in the real space. The WP generation was performed by assigning the displacement
for the atom $i$:
\begin{equation}\label{eqn1}
 \textbf{u}_i=
 A\boldsymbol\epsilon(\textbf{k})
 e^{i\left[\textbf{k}\cdot(\textbf{r}_i-\textbf{r}_0)
 -\omega t\right]}
 e^{-|\textbf{r}_i-\textbf{r}_0|^2/\xi^2}
\end{equation}
where $A$ is the wave amplitude, $\boldsymbol\epsilon(\textbf{k})$ the phonon polarization vector,
$\xi$ the spatial extent (coherence length) of the wave packet, and $\omega$ the eigenfrequency for the wave-vector $\textbf{k}$
of a single branch of the phonon dispersion curve.
The phonon
transmission coefficient $\alpha(\omega,\textbf{k})$ is defined as the ratio between the
energy carried by the transmitted and initial wave packet for a given phonon mode
$(\omega,\textbf{k})$.
The transmission $\alpha_{\text{wp}}(\omega)$ of a WP with a short $\xi\sim\lambda_c$ ($\Delta\omega\sim\omega_c$) is
given by the convolution:
\begin{equation}\label{eqn2}
 \alpha_{\text{wp}}(\omega)
 =\int_0^{\omega_{\text{max}}} \alpha_{\text{pw}}(\omega')
 e^{-\frac{(\omega-\omega')^2}{2\Delta\omega^2}}
 \frac{\text{d}\omega'}{\Delta\omega\sqrt{2\pi}}
\end{equation}
where $e^{-\omega^2/2\Delta\omega^2}$
is a Gaussian WP with $\Delta\omega$ FWHM, and $\alpha_{\text{pw}}$ is the transmission coefficient for
plane waves.
We have confirmed that the capacitor phonon modes correspond to the antiresonance modes of
the PIM, as shown in Fig.~\ref{fig3}(a) and (b).
Therefore, the phonons in the capacitor initially excited by external thermal or optical pumping will
experience spectral narrowing and concentrate to the transverse and longitudinal
modes, at $\omega_1$ and $\omega_2$ respectively, since the interference mirrors totally
reflect the antiresonance-mode phonons but allow the other phonons to transmit.
The two-color regime of the capacitor originates from the distinct dependance of the transmission
coefficient $\alpha(\omega,\theta,\textbf{k})$ on the WP incident angle $\theta$
for the two phonon polarizations $\textbf{k}$, at their respective resonances $\omega =\omega_1,\omega_2$,
as shown in Fig. \ref{fig3}(c).
By modeling the oblique incidence of the WP (with $\theta<\frac{\pi}{2}$), we found that, for the same $\theta$,
$\alpha_T(\omega_2,\theta)>\alpha_L(\omega_1,\theta)$, indicating that
the resonant transverse phonon has a higher transmittance than the longitudinal counterpart through the mirror.
Therefore, for a large aspect ratio ($p\approx1$), the transverse modes with relatively large incident angles
are more susceptible to transmit through the mirrors. Hence after the cooling process, only the longitudinal modes
remain in the capacitor.
While for a small aspect ratio, both the transverse and longitudinal resonance modes can be confined.

The confining effect of an acoustic mode could also be found in a sub-THz acoustic
nanocavity\cite{trigo2002,huynh2006} composed from a spacer held between two Bragg reflectors (BRs).
The defect mode in the nanocavity corresponds to a Fabry-P\'erot (FP) resonance in an inhibited background of the
BR's acoustic bandgap\cite{lanzil2010}. This FP mode can give rise to a total transmission peak
lying inside the wide phonon bandgap\cite{trigo2002}, which can be identified as ``resonance tunneling''.
While the frequency of the FP mode inside the bandgap is well defined, its wavevector $k$
is a complex number, making it difficult to define the wavelength $\lambda$ of the evanescent cavity mode.
The nanocapacitor phonons are, in contrast, well defined with the wavevector in the phonon passband
thus enabling coherent phonon emission from the capacitor.
The BRs are composed of superlattices
while interference mirrors in the phonon capacitor require only single atomic planes,
which enables potential realization of the ultra-compact nanophononic devices.

\begin{figure*}
\includegraphics[width=16.0cm]{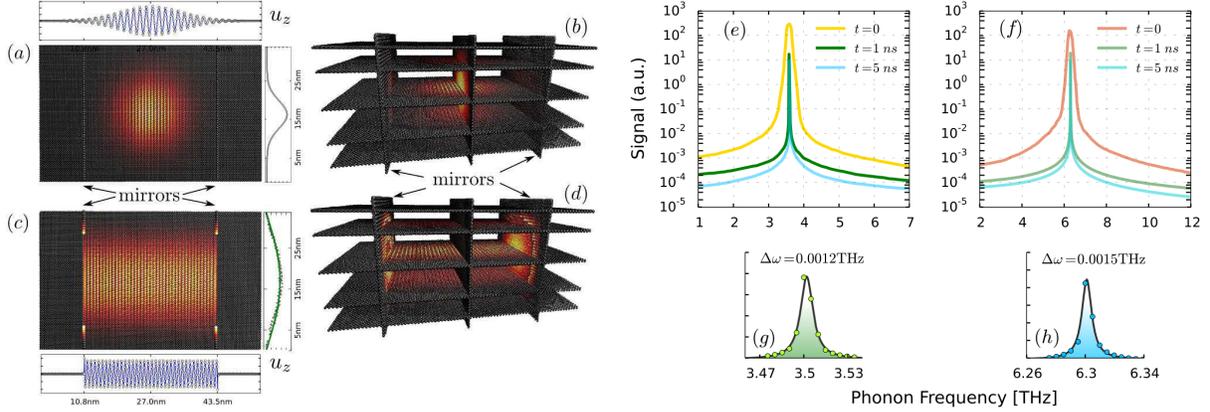}
\caption{\label{fig4}
\textbf{Atomic kinetic energy $E_k$ in the nanocapacitor.}
\textbf{a,b,} Initial 3D phonon wave packet in a cutting plane (\textbf{a}) and in 3D nanocapacitor (\textbf{b}).
\textbf{c,d,}  The final coherent standing-wave phonons with well defined wave-vectors $\pm \textbf{k}$ and wavelengths $\lambda_c$
confined in the nanocapacitor in a cutting plane (\textbf{c}) and in 3D structure (\textbf{d}).
Here several cutting planes in the capacitor are shown to allow for an internal view of phonon energy distribution. Lighter area represents higher $E_k$.
Panels show the atomic displacements along the $x$ and  $z$ directions
of the initial WPs (\textbf{a,b}) and final standing waves (\textbf{c,d}).
Dynamical evolution of both transverse and longitudinal phonon WPs in the nanocapacitor is shown.
\textbf{e,f,} Power spectral density of atomic kinetic energy in the nanocapacitor as a function of time delay $t$
after the launch of the transverse (\textbf{e}) and longitudinal (\textbf{f}) phonon WP.
The peak centered at $\omega_{1,2}$ demonstrates a continuous narrowing as a function of $t$.
At $t=5$ns, the widths narrow down to $\Delta\omega_1=1.2\times 10^{-3}$ THz and $\Delta\omega_2=1.5\times 10^{-3}$ THz,
which correspond to capacitor $Q$ factors of $Q_1=2916$ and $Q_2=4208$, respectively.
\textbf{g,h,} Gaussian fit of the capacitor phonon peak with the measured full width at half-maximum (FWHM) of the  transverse (\textbf{g})
and longitudinal (\textbf{h}) modes.
}
\end{figure*}

To understand better the generation mechanism of the phonon capacitor, we investigated the vibrational
dynamics in the nanocapacitor under a local excitation in the form of a phonon WP
(see Supplementary Material).
We demonstrate through the mapping of the atomic kinetic energy $E_k$ in the capacitor,
shown in Fig.~\ref{fig4}(b) and (d),
that an initial phonon WP centered at ($k,\omega$) finally transforms to a standing wave composed of
two quasi plane waves defined by opposite wave-vectors $\pm \textbf{k}$ and the same frequency $\omega$ in the normal to the mirrors direction. Between the open boundaries of the capacitor, a standing wave with the envelope wavelength of
$\lambda=1/2L_x^c$ comes into existence.
All the other frequency components leak out from the cavity and are absorbed by viscous material at the two ends of the capacitor. In practice, the pump-probe optical technique \cite{huynh2006,lanzil2010} can be used to charge the phonon capacitor, and the considered WP excitation mimics this technique.

\begin{figure}
\includegraphics[width=8.1cm]{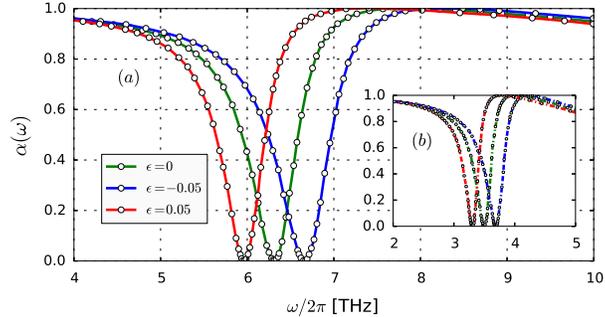}
\centering
\caption{\label{fig5}
\textbf{Shift of interference antiresonance spectral loci caused by tunable strain
$\epsilon$.}
\textbf{a,b,} Antiresonance frequency shifts for longitudinal  (\textbf{a}) and transverse  (\textbf{b}) plane lattice waves, normally incident on interference mirror. The strain is applied in the $\langle 100 \rangle$ direction to the phonon capacitor.
Negative strain $\epsilon<0$, corresponding to a compression on the capacitor,
results in a blue shift of mirror's
antiresonance frequencies, in accordance with positiveness of the Gr\"{u}neisen parameter.
In a similar manner, positive strain red-shifts antiresonance frequencies.
}
\end{figure}

We now turn to the controllable emission of the coherent phonons in the capacitor.
Figure~\ref{fig5} shows the shift of the interference antiresonance spectral loci for transverse and longitudinal phonons caused by
the uniaxial strain $\epsilon$, applied in the $\langle 100 \rangle$ direction to the phonon capacitor.
The strain produces the
change in the local force constants
between Ge and Si atoms in the mirrors, which shifts the antiresonance frequencies $\omega_{1,2}$ of the nanocapacitor, according to the value of the  Gr\"{u}neisen parameter of Si.
Through such a mechanism, we are able to emit coherent phonons and therefore discharge the capacitor by
applying tunable reversible stress at the tips of the device (and the applied strain can be rather small, $\epsilon=1$\%, see Supplementary Material).
The stored phonons are emitted from the capacitor with the group velocity $v_g(\omega,k)=\partial\omega/\partial k$
of the phonon mode.
Once the external stress is released, the phonon emission is switched off due to the recovery of the
capacitor mirrors back to the original antiresonance frequencies.
Therefore, the directional and coherent phonon emission from the nanocapacitor, which can be considered as
``phonon lasing''\cite{vahala2009,mahboob2013}, is flexibly switched by the external stress. According to figure 2, the $two$-$color$ phonon lasing of both transverse and longitudinal coherent lattice waves can be realized with the dual-mode nanocapacitor.
The swift switching of the nanocapacitor enables promising applications in phonon computing\cite{li2012}
and nanoscale memories\cite{jang2008}.

In conclusion,
we introduce a novel ultra-compact nanocapacitor of coherent phonons formed by high-finesse interference mirrors
based on atomic-scale semiconductor metamaterials.
Through MD simulations we show that
the nanocapacitor stores THz monochromatic lattice waves, which can be used for phonon lasing - the emission of coherent phonons.
Either one- or two-color phonon lasing can be realized  depending on the geometry of the nanodevice.
Phonon nanocapacitor can be charged by cooling the sample equilibrated at room temperature or by the pump-probe technique.
The nanocapacitor can be discharged by applying tunable reversible strain, resulting in the emission of
coherent THz acoustic beams.

%
%
%
%

\end{document}